# Magnetic field asymmetry of nonlinear transport in carbon nanotubes

J. Wei, M. Shimogawa, Z. Wang, I. Radu, R. Dormaier, and D.H. Cobden

*Department of Physics, University of Washington, Seattle WA 98195-1560*

We demonstrate that nonlinear transport through a two-terminal nanoscale sample is not symmetric in magnetic field $B$. More specifically, we have measured the lowest order $B$-asymmetric terms in single-walled carbon nanotubes. Theoretically, the size of these terms can be used to infer both the strength of electron-electron interactions and the handedness of the nanotube. Consistent with theory, we find that at high temperatures the $B$-linear term is small and has a constant sign independent of Fermi energy, while at low temperatures it develops mesoscopic fluctuations. We also find significant magnetoresistance of nanotubes in the metallic regime which is unexplained.

The conductance $G$ of a two-terminal sample in linear response must be an even function of applied magnetic field $\mathbf{B}$, that is, $G(B) = G(-B)$ [1,2]. The underlying principle that leads to this Onsager symmetry is the time-reversal symmetry of the equilibrium microscopic dynamics, combined with the fact that $\mathbf{B}$ changes sign on time reversal. There is no such strong symmetry requirement for nonlinear response. Nevertheless, some useful observations may also be made about the nonlinear transport coefficients. These are most readily framed by expanding the current $I$ in powers of the voltage $V$:

$$I = G(B)V + \chi(B)V^2 + \dots , \qquad (1)$$

and focusing on the first nonlinear coefficient $\chi$, which we expand in powers of $B$:

$$\chi(B) = \chi_0 + \alpha B + \dots . \qquad (2)$$

One observation is that for a sample with helical symmetry the sign of the coefficient $\alpha$ in Eq. 2 depends on the handedness [3,4]. This is essentially because the axial vector $\mathbf{B}$ combined with the helicity defines a direction which is inverted if $\mathbf{B}$ is inverted, in the same way that the direction of motion of a screw depends on the sense in which it is rotated. This could for example in principle allow one to distinguish between left- and right-handed chiral carbon nanotubes [5].

A second observation, made only recently [6,7,8], is that the magnitude of $\alpha$ is proportional to the strength of electron-electron (e-e) interactions in the sample, and it can thus in principle be used to deduce the interaction strength. This is true at both high and low temperature $T$. In the high-$T$ limit $\alpha$ can be calculated using a Boltzmann equation approach [6], and it is found to be proportional to the e-e scattering rate, or to $\beta^2$, where $\beta$ is the interaction parameter. In the low-$T$ (mesoscopic) limit it can be calculated either by general diagrammatic techniques [7,9] or within the

Landauer picture of single particle scattering from a self-consistent potential [10,8], and it is found that $\alpha \propto \beta$. The Landauer picture affords a simple understanding of this result, as follows. The Onsager symmetry is obeyed at each energy. In the absence of interactions the total current is the sum over contributions at all energies, and is thus also even in $B$, and so $\alpha$ is zero. However, the electric field due to the applied voltage $V$ induces changes in the local current and electron densities which contain $B$-odd components (as happens for example in the Hall effect). If there are interactions these density changes produce $B$-odd components in the scattering potential, and therefore in $\chi$, which are proportional to $\beta$.

It is well known that e-e interactions are important in single-walled carbon nanotubes because of their one-dimensional (1D) electronic dispersion [11]. Describing the conduction electrons in an infinite clean nanotube as a Luttinger liquid [12,13] allows one to explain the power-law energy dependences of tunneling rates seen in several transport experiments [14,15]. However, real nanotubes are finite in length and often disordered, and the nature of transport in them at high and low $T$ remains an open question. For this reason, we have chosen them as a test system in which to carry out the first specific and detailed measurements of nonlinear coefficients $\chi$ and $\alpha$. We exploit the fact that in a nanotube, unlike a pure 1D system, there is a simple mechanism for generating magnetotransport effects: the dispersion is modified by a magnetic field along the tube axis due to the Aharonov-Bohm phase [16,6]. Our results are in agreement with the general expectations of the theory: near room temperature $\alpha$ is small and its sign is independent of gate voltage, whereas as $T$ is decreased $\alpha$ develops random mesoscopic fluctuations. In addition, we find magnetoresistance in nanotubes in the metallic regime persisting up to room temperature. Our results suggest that basic theoretical questions about magnetotransport in a 1D electron system remain to be addressed.

Each device consists of an individual single-walled carbon nanotube formed by chemical vapor growth from iron catalyst particles [17,18], with two gold contacts patterned by thermal evaporation through a stencil. The substrate is 450 nm of thermal $SiO_2$ on a highly n-doped silicon wafer, to which a gate voltage $V_g$ is applied through a 10 MΩ resistor. An atomic force microscope image of a device (Device 1) containing a nanotube of diameter $d \sim 1.3$ nm and length $L = 4$ μm between the contacts is shown in Fig. 1a, along with the arrangement used to measure $G$ and $\chi$. A sinusoidal bias of rms amplitude $V_0$ at frequency $f$ (650 Hz) is applied to one contact with the other connected to a virtual-earth current preamplifier (Ithaco 1211). A 10 μF capacitor in series enforces zero dc current. The rms harmonic current components $I_f$ and $I_{2f}$ are extracted using lockin amplifiers. In all the measurements $V_0$ is kept sufficiently small ($\sim kT/e$) to ensure that $I_{2f} << I_f$ and that $I_{2f} \propto V_0^2$, so that harmonics beyond $I_{2f}$ are negligible. The first two coefficients in Eq. (1) can then be obtained as $G = I_f/V_0$ and $\chi = 2I_{2f}/V_0^2$.

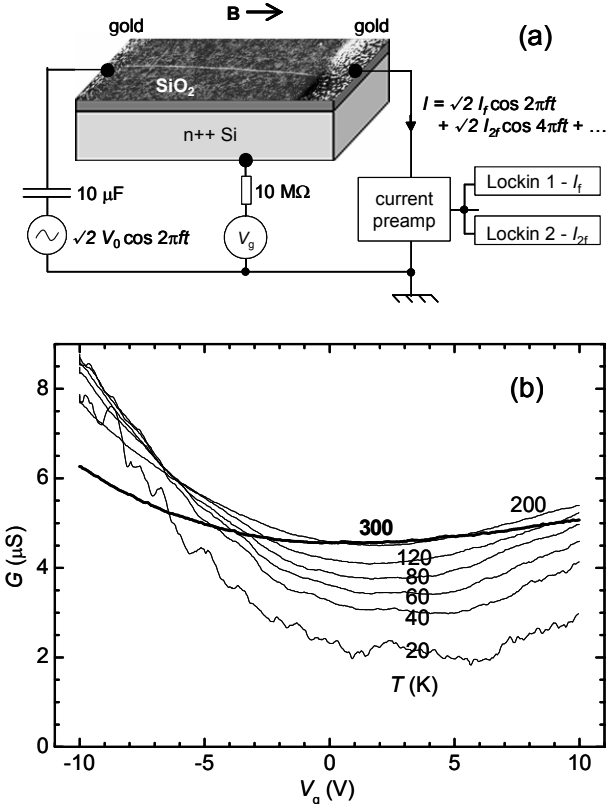

Figure 1. (a) Tapping-mode atomic force microscope image of Device 1 combined with a schematic diagram of the measurement setup. The separation of the gold contacts is 4 μm. The orientation of the magnetic field parallel to the nanotube is indicated. (b) Linear conductance vs gate voltage at a series of temperatures.

Before discussing the nonlinear behavior, let us see what can be understood of the linear conductance.

The linear behavior of Device 1 at $B = 0$ is illustrated in Fig. 1b. The weak dependence on $V_g$ at room temperature is characteristic of a metallic (or possibly a small-gap) nanotube. We have also made detailed measurements on a semiconducting nanotube, but the strong $V$ and $T$ dependences arising from bandgap effects make interpreting the nonlinear behavior more difficult in them. In Device 1, as $T$ is decreased from 300 K, $G$ rises and passes through a maximum at a temperature which depends on $V_g$ before falling again. For $T < 20$ K, a dense 'grass' of aperiodic and unreproducible Coulomb blockade oscillations appears (not shown), and by $T = 4.2$ K, $G$ is too small to measure.

The negative value of $dG/dT$ at room temperature indicates that the resistance is not dominated by the contacts, since poor contacts would ensure positive $dG/dT$ [14]. We also know from characterising many similar devices that our contacts are reliable and have high transparency, ie, the contact resistance $R_c$ is not much larger than the ideal value of $h/4e^2$. If we assume the additional resistance is distributed along the nanotube, then in the high-$T$ limit it can be characterised by a backscattering length $l_b$ given by $G^{-1} = R_c + (h/4e^2)L/l_b$ [19]. At 300 K, $G \sim 5$ μS, and we find $l_b > \sim 150$ nm. This is shorter than the phonon scattering length which is approximately 1 to 2 μm [20], implying that backscattering is predominantly due to disorder, although some phonon scattering can explain the negative $dG/dT$ at room $T$. This is consistent with the behavior in the low $T$ regime where from the Coulomb blockade we infer that the nanotube breaks up electrically into a series of submicron regions [21]. We do not know the precise origin of the disorder, which is much higher than in the cleanest nanotube devices [20]. It may be explained by contaminants on the nanotube's surface associated with our growth process.

The effects of magnetic field on the linear conductance are illustrated in Fig. 2. At room $T$, $G$ decreases approximately quadratically with $B$ up to ±16 T. Fitting it to $G(B) = (1+\gamma B^2)G_0$ at each gate voltage, we find that the parameter $\gamma$ varies steadily from about $-2\times10^{-4}$ T$^{-2}$ at $V_g = 0$ to $-4\times10^{-4}$ T$^{-2}$ at $V_g = 5$ V. In the semiconducting device at 200 K we found $\gamma$ to be *positive*, reaching a peak of $+4\times10^{-3}$ T$^{-2}$ close to threshold but maintaining a value of $+2\times10^{-4}$ T$^{-2}$ in the metallic regime. A positive magnetoconductance near threshold in a semiconducting nanotube can be explained by a decreasing bandgap [6,16,22,23]. (The contribution of the gold leads to the resistance is negligible.) However, to our knowledge no mechanism has been put forward to explain significant magnetoconductance in the metallic regime .

As $T$ decreases, $G$ develops nonperiodic oscillations as a function both of $V_g$ (see Fig. 1b) and of $B$ (see Fig.

2a). Fig. 2b is a greyscale plot of $G(V_g,B)$ at 20 K. Fig. 2c is a histogram of the $B$-symmetrized and antisymmetrized parts of $G$ at 20 K, showing that the antisymmetric part is much smaller than the symmetric part. Since the absence of an antisymmetric component is required by Onsager symmetry, this is evidence that the measurement is indeed effectively two-terminal. (The visible deviation from symmetry about $B = 0$ in the 20 K (dashed) sweep in Fig. 2a, and the broadening of the black peak in Fig. 2c, resulted from drift over the two-hour timescale of the magnetic field sweep.)

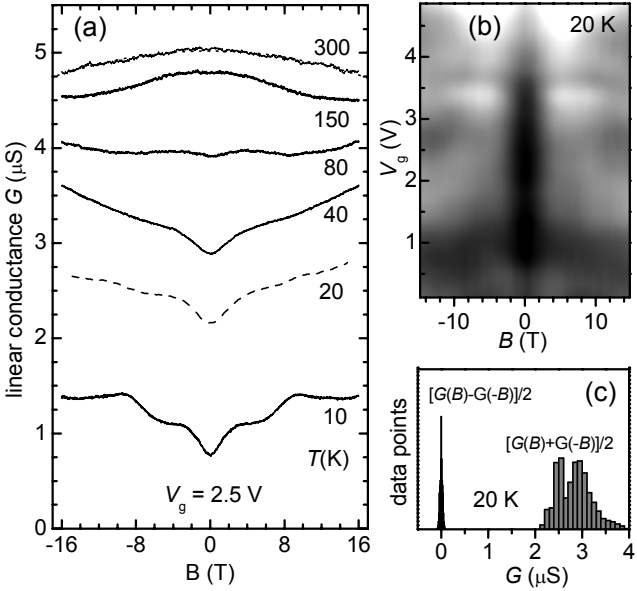

Figure 2. (a) Linear conductance $G$ vs. magnetic field $B$ at a series of temperatures for a fixed gate voltage, measured using the setup in Fig. 1. (b) Grayscale plot of $G(V_g,B)$ at $T = 20$ K. Black is 2.2 μS, white is 3.5 μS. (c) Histogram of $B$-symmetrized and $B$-antisymmetrized parts of $G$, averaged over all $V_g$ and $B$, from the dataset in (b).

Oscillations as a function of $V_g$ and $B$ are predicted by a single-particle model of quantum interference [23,24] in the presence of multiple scatterers. In this model, one anticipates a characteristic magnetic field period $B_c \sim 4(h/e)/(L_{eff}d)$, corresponding to the change in magnetic flux which alters the phase difference between typical electron paths in the $K$ and $K'$ subbands by $2\pi$. Estimating the effective path length $L_{eff}$ to be the lesser of the nanotube length $L$ and the thermal length $L_T \sim h v_F/k_B T$, where $v_F = 8\times10^5$ ms$^{-1}$ is the Fermi velocity, for $T = 20$ K we obtain $L_{eff} \sim L_T \sim 2$ μm and $B_c \sim 6$ T. This is compatible with the oscillations seen in Fig. 2a. The model also predicts a gate voltage oscillation period $\sim h v_F/(e L_{eff})$ of a few mV (taking into account that the capacitance is dominated by the gate). Such short-period oscillations would not have been resolved in these measurements. However, it is clear that there are features in Fig. 2b which vary much more slowly with $V_g$. In particular, at 20 K there is a dip at $B = 0$ with a half-width of ~2 T which persists over the entire range of $V_g$. Again, no theory is available, though this behavior bears a suggestive resemblance to the weak localization seen in conventional metals and recently in multiwalled nanotubes [25].

We are now ready to consider the behavior of the nonlinear coefficient, $\chi$. Its dependence on $B$, $V_g$ and $T$ is illustrated in Fig. 3. Figs. 3a-c are greyscale plots of $\chi(V_g,B)$ at three different temperatures. Figs. 3d and 3e show line traces of $\chi$ vs $B$ at two selected gate voltages. At the highest temperature of $T = 200$ K (bold traces), $\chi$ is small and varies slowly with $B$ up to ±16 T. Like $G$, $\chi$ develops oscillations as a function of both $V_g$ and $B$ as the sample is cooled. In contrast with $G$ however, $\chi$ is not symmetric about $B = 0$. Figs. 3f and 3g are greyscale plots of the symmetric and antisymmetric parts of $\chi$ at $T = 20$ K.

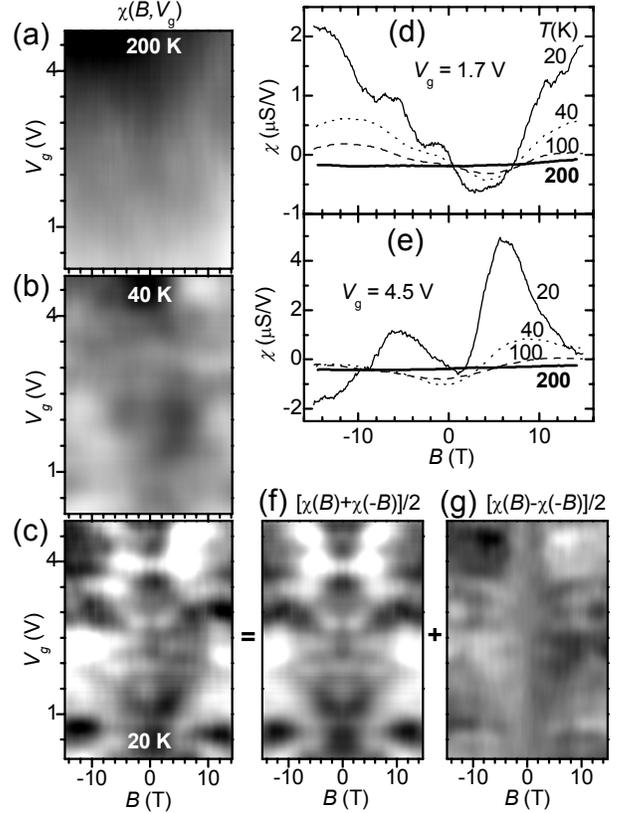

Figure 3. (a) Variation of the $V^2B$ coefficient $\alpha$ with $V_g$ at a series of temperatures. Inset: log-log plot of $\langle\alpha^2\rangle$ vs $T$, where the straight line indicates a $1/T^2$ dependence.

We note that $\chi$ depends on both the intrinsic asymmetry of the device and the asymmetry of the measurement (Fig. 1a): applying a bias $V_0$ to the source generates a nonlinear 'self-gating' current $I_{2f,sg} \sim (dG/dV_g)V_0^2/4$ due to the change in $G$ caused by the resulting change in average potential difference between gate and nanotube [26]. This current contribution must, like $G$, be symmetric in $B$ and cannot contribute to the $B$-antisymmetric part of $\chi$ reported above.

To date there exist no predictions specific to nanotubes with which we can quantitatively compare these measurements of $\chi$. Nevertheless, the theory mentioned in the introduction leads to qualitative expectations for the behavior of the $B$-linear coefficient $\alpha$ in Eq. (2). At high $T$, the sign of $\alpha$ should depend on the handedness of the nanotube [5,6]. In this regime $\alpha$ should vary slowly, without oscillating, as a function of Fermi energy and thus of $V_g$. At low $T$, in the mesoscopic regime, as a result of disorder one expects mesoscopic fluctuations of $\alpha$ characterised by correlation functions [7]. Since the disorder should have no preferred chirality, one expects $<\alpha> = 0$, where the average is taken over disorder realizations. In addition, Ref. [7] predicts $\langle \alpha^2 \rangle \propto \beta^2/T^2$ for a normal mesoscopic metallic sample.

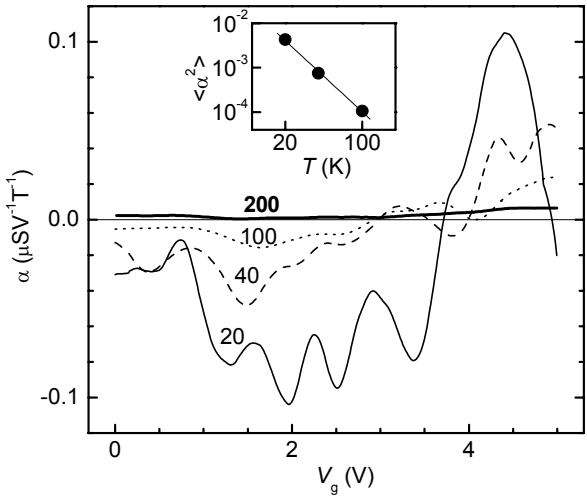

Figure 4. (a) Variation of the $V^2B$ coefficient $\alpha$ with $V_g$ at a series of temperatures. Inset: log-log plot of $<\alpha^2>$ vs $T$, where the straight line indicates a $1/T^2$ dependence.

We extracted values of $\alpha$ from the data by fitting a straight line of the form $\chi_0 + \alpha B$ to the data points of $\chi$ vs $B$ in the range $-2\ T < B < +2\ T$, doing so at each value of $V_g$ and $T$. The results for $\alpha$ are shown in Fig. 4. At the highest temperature (200 K), $\alpha$ is small and varies slowly with $V_g$ without changing sign. This is consistent with the above expectations. As $T$ decreases $\alpha$ develops oscillations which cause its sign to alternate as a function of $V_g$, again consistent with the expectations. In the inset we plot $\langle \alpha^2 \rangle$, obtained by averaging $\alpha^2$ over $V_g$, against $T$. The results are surprisingly consistent with the $1/T^2$ dependence (solid line) mentioned above, in spite of the ostensible inapplicability of the calculation in Ref. [7].

In summary, we have carried out the first experimental study of a new transport coefficient in nanoscale devices, namely the magnitude of the $V^2B$ term in the $I$-$V$ characteristics. This coefficient provides a way to quantify the electron-electron interaction strength, which is of particular interest in our chosen system of single-walled carbon nanotubes. We also find unexplained magnetoresistance in disordered metallic nanotubes at high and low temperatures that acts as a further indication that basic aspects of these 1D conductors remain to be addressed theoretically.

We thank A. Andreev, E. Deyo, B. Spivak, O. Vilches and N. Wilson for useful discussions and J. Chen for providing the catalyst particles. This work was partly supported in part by the UW Royalty Research Fund, an NSF IGERT fellowship (M.S.), a UW UIF fellowship (I.R.), and a Mary Gates fellowship (R.D.)